\documentclass[12pt]{article}
\usepackage[a4paper, margin=25mm]{geometry} 

\usepackage{fancyhdr}                       
\usepackage{indentfirst}                    
\usepackage{graphicx}
\usepackage{cite}                           
\usepackage{wrapfig,rotating}
\usepackage{amssymb,amsmath,array}
\usepackage{authblk}

\newcommand{\JxB}{$\rm{\textbf{J}}\!\times\!\rm{\textbf{B}}$ }
\begin{document}
\pagestyle{fancy}                           

\title{Liquid Metal Flow Controls at Liquid Metal Experiment}
\author[1,2]{M. Modestov}
\author[2,3]{E. Kolemen}
\author[3]{E. P. Gilson}
\author[2]{J. A. Hinojosa}
\author[3,4]{H. Ji}
\author[3]{T. Kunugi}
\author[3]{R. Majeski}
\affil[1]{Nordita, KTH Royal Institute of Technology and Stockholm University, 10691, Stockholm, Sweden}
\affil[2]{Department of Mechanical and Aerospace Engineering, Princeton University, Princeton NJ 08544, USA}
\affil[3]{Princeton Plasma Physics Laboratory, Princeton, NJ 08543, USA}
\affil[4]{Department of Astrophysical Sciences, Princeton University, Princeton NJ 08544, USA}
\maketitle

\begin{abstract}
Liquid metal flow behavior under magnetic field and electric current is investigated in experiment and numerical simulations. Several aspects of the resulted Lorentz force action are discussed and demonstrated. The enhanced flow mixing induced by the non-uniform current density appears to be crucial for the heat transfer efficiency. Also the outflow heat flux is strongly affected by the action of the \JxB force.
\end{abstract}

\section{Introduction}
Despite considerable progress in magnetic fusion studies several fundamental phenomena still hinder it from profitable efficiency. One of the main problems is the first wall disruption under severe heat flux from the hot plasma.
A flowing layer of a liquid metal (LM) as an alternative tokamak interior wall reveals growing recognition \cite{Morley_2004,Jaworski_2013}. Its main idea as a heat removal instrument is straightforward, however many basic and engineering properties require much efforts in this field.
For these purposes in PPPL the Liquid Metal Experiment (LMX) has been designed in a form of a rectangular channel allowing investigation of the heat transfer in LM under various conditions \cite{Rhoads_2014}.
Plans for a fully toroidal experiment to study the flow under magnetic fields and electric currents similar to tokamak divertor operation are being developed.

Real experiment data is of high value, however it yields the whole picture of net influence of all possible factors, concealing role of basic physical mechanisms. In this sense proper MHD simulation reproducing the experimental results is crucial for understanding undelaying physics \cite{Morley_2004,Kirillov_1995,Krasnov_2012}.
This will also enable extrapolation of the present results to the expected behavior in fusion reactor and the development of improved designs and control systems.

For successful operation of LM divertor numerous aspects must be thoroughly studied.
The whole problem involves coupling between flow dynamics, heat transfer, electromagnetic interaction with conducting media, as well as free-surface issues.
In this work we are mostly interested in the heat spreading in the LM flow under magnetic field and electric current running through the LM.
We discuss the experimental setup and multiple effects of the resulted Lorentz force.
For computer modeling we imply certain simplifications, however the numerical setup is very close to the LMX parameters.
In the following sections we describe this problem in more details and compare simulation results with the experimental measurements.

\section{Governing Equations and Numerical Setup}

This paper focuses on numerical simulations of the LM flow.
In particular we are interested in the electric current running through the flowing LM placed in external magnetic field and the influence of resulted Lorentz force upon the heat spreading and general behavior of the LM flow.
The LM is described as an incompressible conducting fluid, accounting for its viscosity and thermal properties. The governing equations are the following:
\begin{equation}
\begin{split}
\label{sys0}
 \rho\frac{\partial \textbf{u}}{\partial t} + \rho (\textbf{u}\cdot\nabla)\textbf{u} +\nabla p& =
  \mu \nabla\left[\nabla \textbf{u} + (\nabla \textbf{u})^T \right] + \left[ \textbf{J}\times\textbf{B} \right] \\
 \frac{\partial T}{\partial t} + \textbf{u} \nabla T &= \alpha \triangle T \\
 \nabla \textbf{u} &= 0 \\
 \textbf{J}&=\sigma \left(-\nabla V + \textbf{u}\times\textbf{B}\right)
\end{split},
\end{equation}
where $\rho$ is the LM density, $\textbf{u}$ is the flow velocity, $p$ is the hydrodynamic pressure, $\mu$ is the dynamic viscosity, $\textbf{J}$ is the total current density, $\textbf{B}$ is the magnetic field, $T$ is the temperature, $\alpha$ and $\sigma$ are the thermal diffusivity and the electric conductivity, respectively, $V$ is the electric potential.
Equations (\ref{sys0}) have been simulated with the help of the COMSOL Multiphysics software package. The solver is based on the finite element approach aiming for easy coupling between different physics. It should be mentioned that the problem is essentially three dimensional, as the magnetic field, the current density and their resulting Lorentz force are orthogonal to each other, as shown in Fig.~1.

\begin{figure}
 \includegraphics[height=5cm]{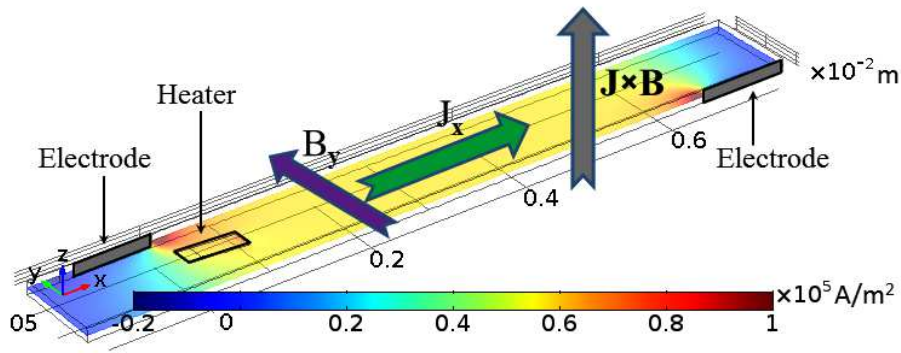}
 \caption{Numerical setup together with the typical current density distribution.}
 \label{fig1}
\end{figure}

In our modeling the fluid density is assumed to be constant, neglecting LM thermal expansion and omitting natural convection phenomenon. In fusion reactor these effects must be taken into account, however under LMX conditions temperature varies within $\sim\!30$K which is not enough for noticeable density change.
Besides the heater position at the top of the flow diminishes the role of convection as well. 
The Joule heating has also been omitted from the model, as it produces much smaller temperature rise than the heater. 
The magnetic field is assumed to be constant in time and space, namely $\textbf{B}=(0,B_y,0)$. For the larger part of the flow the field variation is negligible, however in experiment the magnet size is shorter than the channel length, so that there are side affects.
Finally the LMX represents a free-surface flow, which under relatively strong Lorentz force leads to unstable and wavy flow surface. In the simulation setup we have a flat surface for the top wall with slip boundary condition.
Other boundaries are set as non-slip except for constant velocity at the inlet (left channel side) and outlet boundary condition at the right channel side; at the inlet the temperature is set to a certain fixed value as well.
The heater surface facing the channel provides heat flux with constant power supply, so that the heater temperature depends on its surface area and the flow velocity; other walls are thermal insulators.
Also in the experiment the heater is held at slight angle $\sim\!4^{\circ}$ with respect to the horizontal flow surface to ensure good thermal contact at different flow velocity and height. In the simulation we use similar setup with the inclined heater being partially immersed in the LM, so that the different flow height yields corresponding changes in the thermal contact area. The electric current density is computed due to the voltage difference set at the electrodes (shown in Fig.~1); the other walls are set as insulators as the experimental channel has an insulating acrylic liner.
For diagnostics the experimental setup provides measurements of the flow rate and height together with the temperature at the bottom of the channel due to numerous thermocouples.
Hence our simulation setup is almost analogous to the real experimental one, taking into account actual galinstan properties, used in LMX \cite{Morley_2008-Galinstan}.

\section{Lorentz Force Effects in LMX}
One of the major purposes of the current LMX version is to study the impact of the Lorentz force on the heat transfer in LM as well as its general influence on the flow.
From our preliminary simulations, experimental data and theoretical expectations it has been found that the Lorentz force affects the flow in manifold ways.
First of all the \JxB term is aligned with the \textit{z}-axis, acting as effective gravity, which, under free-surface flow, modifies the flow height and velocity.
This, in turn, changes the heater contact area with the LM.
The experimental setup magnet envelops most of the channel, however there are two minor parts of the channel outside the magnet.
It means that the actual gravity is different in different parts of the channel creating corresponding jumps in LM height. It results in unstable and wavy flow surface which may cause additional mixing.
Another issue deals with the specific location of the electrodes at the sides of the channel.
As shown in Fig.~1 such a configuration yields non-uniform current density, producing huge peaks near the electrodes edges. It generates localized secondary flows enhancing LM mixing under the heater and closer to the outlet region.
The magnetic field itself stands for the MHD drag and it modifies the boundary layer thickness as well.

Each of these aspects deserves separate detailed investigation.
Below we present our simulation results and find out which of these factors prevails in heat conduction.

\subsection{Lorentz Force as Effective Gravity}
We start our numerical investigations with a simpler but fundamental problem studying flow height influence upon the heat conduction properties. It is motivated by the fact that in experiments significant flow height change has been observed due to \JxB force. The latter can be considered as effective gravity and under typical experimental conditions with $B_y\!\approx\!0.3$ $\rm{T}$ and $J_x\!\approx\!10^5$ $\rm{A/m^2}$ the resulting force reaches half of the Earth gravity magnitude. Hence the actual gravity varies from $\approx\!0.5g$ to $\approx\!1.5g$, where $g\!=\!9.8$~$m/s^2$ is the standard gravity.
For a free-surface flow gravity variation leads to corresponding change in height which can be estimated in a similar way as hydraulic jump phenomenon. Due to the mass conservation the flow velocity changes with height as well. Each of these two flow parameters has a strong influence upon the heat conduction. In 3D geometry rough analysis yields $h^{-3}$ dependence of the bottom temperature against the flow height and $u^{-2}$ dependence against the flow velocity. At the same time for a fixed flow rate the velocity is inverse proportional to the flow height, $u\sim Q/h$ where $Q$ is the flow rate, so we expect decrease of the bottom temperature with the increase of the flow height.

In LMX the flow height varies within $9-16$ $\rm{mm}$ under $B_y\!\approx\!0.3$ $\rm{T}$ with current altering $J_x\pm10^5$~$\rm{A/m^2}$. In order to study the height effect on the heat conduction we exclude the electro-magnetic physics from the model and perform several simpler simulations at fixed flow rate varying the flow height only, imitating one aspect of \JxB effect.
For this setup the contact area with the heater also remains constant. In Fig. 2a we present the mean bottom temperature at the 20 cm vicinity downwards the heater (such an area has been chosen due to thermocouple array in LMX) and the outlet temperature taken at the bottom center.
Simulations have been performed for two values for the flow rate; the height variation and the flow rate are taken similar to experimental values.
First of all this figure shows temperature decrease with the flow height growth, which is in line with analytical estimates.
The larger flow rate also decreases the bottom temperature, demonstrating dependence on the flow velocity. The difference between the two temperature axes range partially reflects the fact that at the outlet temperature is not uniform, but has a peak at the central part of the channel, while the mean bottom temperature is reduced due to colder areas at the sides of the channel, as shown in Fig. 3. Hence, based on Fig. 2a we may assume that downward \JxB force enhances gravity and decreases the flow height with corresponding growth of the bottom temperature. In the opposite case the upward \JxB force reduces the actual gravity, so that the flow becomes thicker and slower, while the channel bottom gains less heat.

\begin{figure}
 \includegraphics[width=16cm]{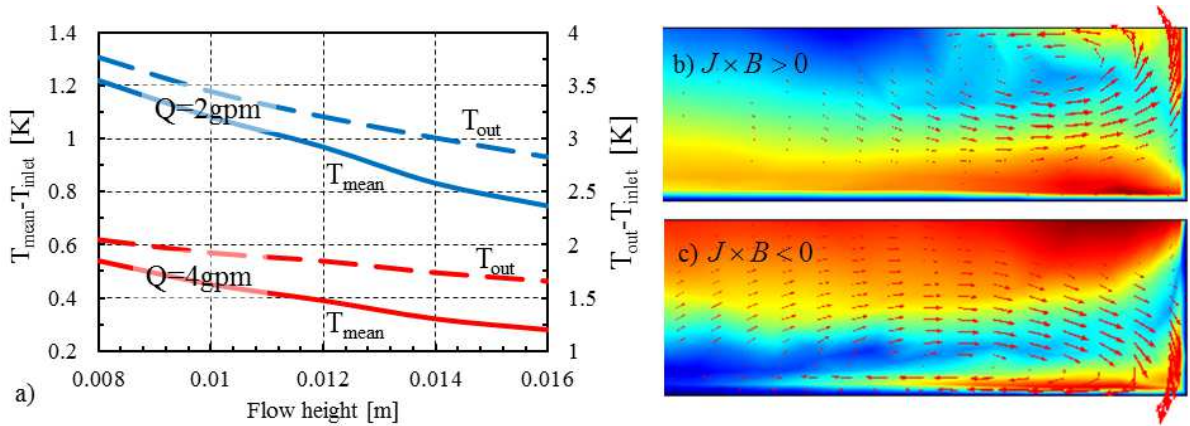}
 \caption{a) Bottom temperature dependence versus the flow height for two flow rate, Q=2gpm and Q=4gpm. Solid lines stand for the mean bottom temperature in 20 cm vicinity after the heater, dashed lines depict the outflow temperature at the in the center of the bottom. b) and c) Velocity field at the y-z cross section near the electrode (at the right side) for two direction of the \JxB force; background stands for the total velocity magnitude, vector field reflects $(u_y;u_z)$ velocity components.}
 \label{fig2}
\end{figure}

\subsection{Non-Uniform Current Density and Secondary Flow}
The present version of the LMX is equipped with two electrodes located at the sides of the channel as shown in Fig. \ref{fig1}.
The current density exhibits uniform distribution in the main part of the channel, which in combination with transverse magnetic field, $B_y$ produces uniform force in vertical direction.
Closer to the electrodes there is a concentration of electric field amplifying current density.
Hence, locally we obtain stronger field directed upwards or downwards depending on the applied voltage sign.
In real free-surface flow it destabilizes the flow surface, however in simulations the top boundary is flat. Nevertheless we can easily see the action of this force by velocity field presented in Fig.~2(b,c) at the cross-section of the channel in $y-z$ plane near one of the electrode edge.
The arrow field shows strong motion in vertical direction near the electrode, which in turn generates a secondary flow across the channel.
The flow patterns are not symmetrical due to different boundary conditions at the bottom (non-slip) and top (slip) boundaries.
Such a local peak of the \JxB force eventually decay as moving from the electrodes to the channel midst, where the force becomes uniform.

It is important to notice that this secondary flow is strong enough to modify the overall flow structure, playing a role of invisible obstacle. As will be demonstrated in the next section it results in the flow divergence near the electrodes, which has also been observed in experiments.
In addition to that the heater is located near one of the electrodes, so that the enhanced mixing is expected to affect the heat transfer efficiency.

\subsection{Most-Completed Numerical Setup}

Finally we perform series of simulations taking into account several effects reflecting more relevant LMX conditions. The flow height is adjusted to the electric current amplitude according to the experimental height measurement.
The heater position is bounded to the channel bottom so its thermal contact area is changing with respect to the height.

\begin{figure}
 \includegraphics[width=16cm]{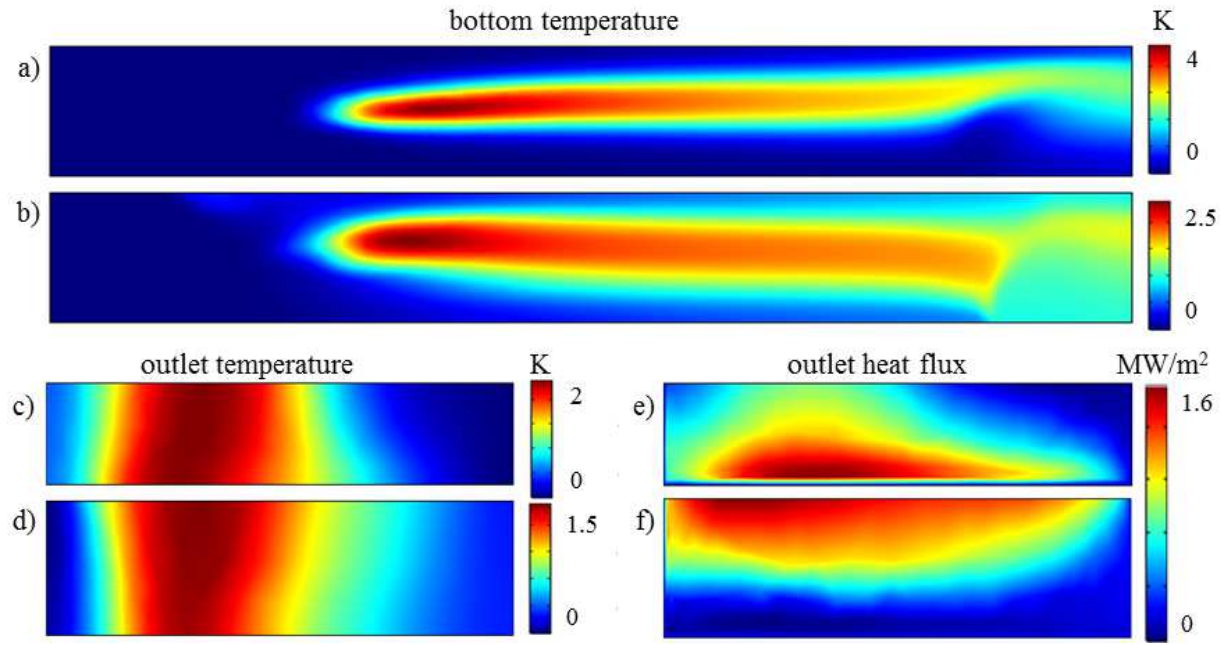}
 \caption{Temperature field at the bottom (a,b) and at the outlet boundary (c,d) together with the outflow heat flux distribution (e,f); panels (a,c,e) are plotted for downwards \JxB force, while cases (b,d,f) correspond to upwards \JxB force.}
 \label{fig4_T}
\end{figure}

In Fig.~\ref{fig4_T} we present the temperature distribution at the bottom (a,b) and the outlet (c,d) for different directions of the \JxB force.
First of all we see noticeable difference between the two cases at the bottom temperature fields.
For the upwards force the higher temperature tail is wider than the one for the downwards force; the maximal temperature also varies as can be seen from the legend panel.
In addition to that we mark essentially non-symmetric heat spreading with respect to the central line of the channel. Closer to the outlet one can observe apparent flow deviation near the electrode edge, where the current density increases significantly.
The outlet temperature distributions look similar, although the upwards (d) case exhibits larger dispersion through the boundary.
In Fig.~\ref{fig4_T}(e,f) we show distribution of the heat flux at the outlet, which reveals huge difference of the two cases. In the downwards setup most of the heat is advected closer to the bottom of the channel while in the upwards case the heat is moving near the top surface. Such distributions reflect corresponding velocity fields, occurring due to the secondary flow.

\begin{figure} 
 \includegraphics[width=16cm]{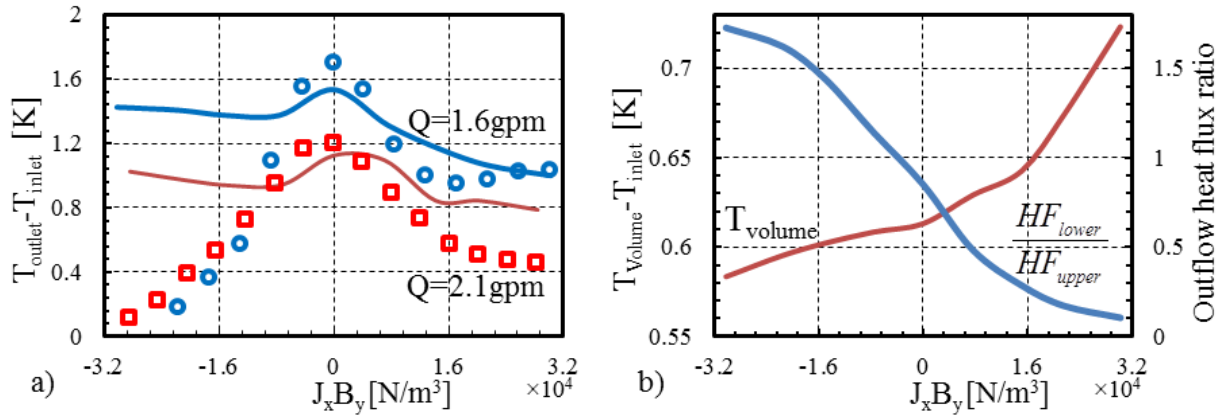}
 \caption{a) Comparison of experimental outflow temperature measurement (markers) with simulation results (solid lines) for two flow rates. b) Increase of the total volume temperature and heat flux ratio ratio at the outlet (lower to upper parts). Both plots are plotted against \JxB force acting in experiment.}
 \label{fig5_final}
\end{figure}

Real experiment involves several interplaying phenomena, such as heat transfer, LM interaction with electric and magnetic fields, proper boundary layer formation and others.
For these reasons together with the above speculations it is really hard to expect good agreement between the simulation results and experimental data.
Still in Fig.~\ref{fig5_final}(a) we compare the outlet temperature obtained from the simulations and provided by LMX. The two sets of points and curves correspond to two flow rates.
Despite noticeable difference between the experimental markers and the simulation curves we should mark close absolute values of the temperature differences. Also both results reveal similar peak of the outlet temperature at small currents. Larger discrepancy at higher currents may indicate influence of the unstable free-surface or other boundary effects which have not been captured in our simulations.
According to Fig. 2 positive \JxB force corresponds to thicker flow and weaker temperature response. Similar trend can be noticed in simulation curves, while the experimental data exhibits different behavior.
Most likely this result indicates crucial role of the flow mixing for the heat conduction rather than variation of the flow thickness.

Few more interesting effects of the Lorentz force are plotted in Fig.~\ref{fig5_final}(b). Along the left axis we present variation of the LM temperature averaged over the whole channel volume.
Absolute variation of the volume temperature is pretty small, however relative change reaches $25-30\%$, which might be important for higher heat loads.
This temperature grow must be attributed to the enhanced heat transfer demonstrated in Fig.~\ref{fig4_T}(a-d).
On the right axis of Fig.~\ref{fig5_final}(b) we plot a ratio of the heat flux through the outlet, $H\!F_{lower}$ corresponds to the averaged heat flux through the lower part of the outlet, $z\leq0.5h$, where $h$ is the flow thickness, $H\!F_{upper}$ is computed for $z\geq0.5h$.
This heat flux ratio reflects corresponding distribution from Fig.~\ref{fig4_T}(e,f).
On one hand the upward \JxB force intensifies mixing, but it also concentrates most of the heat flux in the upper layers of the flow (ratio$<\!1$), which might lead to possible LM evaporation or heat radiation from the surface in real tokamak application. On the other hand the downwards Lorentz force shifts the heat flux to the bottom of the channel, making the heat removal more secure.

\section{Conclusions}
The role of electro-magnetic force in thermal conduction in LM has been studied with the help of experiment and numerical simulations. We have discussed multiple possible effects of the \JxB force and conclude that additional flow mixing induced by the non-uniform current density has the dominate role in the present heat transfer experiments.
Additional investigations are needed to understand the discrepancy in the obtained results as well as to realize the most efficient divertor design for tokamak application. Strong variation of the outflow heat flux due to the Lorentz force action encourages further research in this field. The detail mechanism and optimal current configuration has not been fully understood, although it is clear that the electric current running through the LM can be used to control the uniformity of the heat distribution in conducting fluid.


\vspace*{1cm}
\noindent\textbf{\textit{References}}
\begin{thebibliography}
\bibliographystyle{}
\bibitem{Morley_2004} MORLEY N. B., et al., ``Progress on the modeling of liquid metal, free surface, MHD flows for fusion liquid walls'', Fusion Engineering and Design \textbf{72} (2004) 3.

\bibitem{Jaworski_2013} JAWORSKI M. A., et al., ``Liquid lithium divertor characteristics and plasma-material interactions in NSTX high-performance plasmas'', Nucl. Fusion \textbf{53} (2013) 083023.

\bibitem{Rhoads_2014} RHOADS, J. R., et al., ``Effects of magnetic field on the turbulent wake of a cylinder in free-surface magnetohydrodynamic channel flow'', J. Fluid Mech. \textbf{742} (2014) 446.

\bibitem{Kirillov_1995} KIRILLOV I. R., et al., ``Present understanding of MHD and heat transfer phenomena for liquid metal blankets'', Fusion Engineering and Design \textbf{27} (1995) 553.

\bibitem{Krasnov_2012} KRASNOV D., et al., ``Numerical study of magnetohyrodynamic duct flow at high Reynolds and Hartmann numbers'', J. Fluid Mech. \textbf{704} (2012) 421.
\bibitem{Morley_2008-Galinstan} MORLEY N. B., et al., ``GaInSn usage in the research laboratory'', Rev. Sci. Instrum. \textbf{79} (2008) 056107.
\end{thebibliography}
\end{document}